\documentclass{article}[15pt]      

\usepackage{amsmath,amssymb,epsfig,color,calc,rotating,graphics}
\usepackage{ulem}

\usepackage{float}

\usepackage[pagewise]{lineno}

\usepackage{url}

\usepackage{natbib}
\bibpunct{(}{)}{;}{a}{}{,}

\newcommand{\qed}{\nobreak \ifvmode \relax \else
      \ifdim\lastskip<1.5em \hskip-\lastskip
      \hskip1.5em plus0em minus0.5em \fi \nobreak
      \vrule height0.75em width0.5em depth0.25em\fi}

\begin{document}

\title{Rich dynamical behaviors from a digital reversal operation 
\vspace{0.2in}
\author{
Yannis Almirantis$^1$ and
Wentian Li$^{2,3}$\footnote{Corresponding author (wtli2012@gmail.com; wentian.li@stonybrook.edu)} \\
{\small  1. Theoretical Biology and Computational Genomics Laboratory, Institute of Bioscience and Applications}\\
{\small National Center for Scientific Research ``Demokritos", Athens, Greece}\\
{\small  2. Department of Applied Mathematics and Statistics, Stony Brook University, Stony Brook, NY, USA}\\
{\small  3. The Robert S. Boas Center for Genomics and Human Genetics}\\
{\small  The Feinstein Institutes for Medical Research, Northwell Health, Manhasset, NY, USA}\\
\\
}
\date{\today}      
}  
\maketitle                   
\markboth{\sl }{\sl }

\begin{center}
{\bf ABSTRACT} 
\end{center}

Repeatedly adding or subtracting the digital reversal to or from an integer,
depending on which one is larger, can be treated as a dynamical system. 
On one hand, a three-digit version of this map running only two steps is 
the 1089 mathematical trick problem; on the other hand, this mapping can be 
compared to John Conway's reverse-add-then-sort (RATS) iteration, as well as
the 3x+1 problem, also known as Collatz's map. We numerically run this map and 
find interesting dynamics, including limiting cycles with unusual periodicity 
and length-8 diverging trajectories.

\large

\section*{Introduction}

\indent

The late ``mathematical magician" professor John Conway \citep{houston}
once proposed this iteration or mapping called ``reverse(R), add(A), 
then(T) sort(S)" (RATS) \citep{conway}, with a positive integer at step $i$, $D_i$, 
being mapped to another integer at step $i+1$, $D_{i+1}$, through
a summation of $D_i$ and its digital reversal $rev(D_i)$,
followed by its digital sort:
\begin{equation}
\label{eq-conway}
D_{i+1}= dsort[ D_i+ rev(D_i) ].
\end{equation}
Example of digital reversal:  $rev(1230)= 321$; and example of digital sort: $dsort(8120)=128$.
Keep doing it again and again, or in a phrase used in \citep{conway},
``set numbers in motion", where it will end up to? 
Eq.(\ref{eq-conway}) has 2-cycle limiting dynamics \citep{guy},
if, for example, starting from $D_0=1111122267$.
Higher periodicities dynamics of Eq.(\ref{eq-conway}), such as 3,4,5,6,7,8,9,10,11,12,14,18,24-cycles,
have also been observed  \citep{guy, guy2,cooper1,cooper2}. On the other hand,
if the initial $D_0=12334444$, then $D_1=55667777$, and $D_2=123334444$ has a similar
pattern as $D_0$ but with an extra 3. $D_4$ is again similar to $D_0$ but has
two extra 3. The number $D_4$ and subsequent iterations keep increasing the sequence length and
never settle to a periodic orbit. This is a case of diverging trajectory.

What limiting dynamics does Eq.(\ref{eq-conway}) (RATS) lead to depends
on the initial $D_0$.  One may metaphorically think of a real 
planetary motion governed by Newton's physics. The rule -- here 
Newton's gravitational law -- is fixed, but the time-to-infinity
limiting dynamics still depend on the initial conditions, resulting
in different orbits, including some escaping the Sun's gravitational pull
towards infinity. 

Here we propose another mapping that also has a rich range of dynamical
behaviors as follows:
\begin{eqnarray}
\label{eq-map}
D_{i+1} =
\left\{
\begin{matrix}
D_i - rev(D_i) & \mbox{if $rev(D_i) < D_i$} \\
D_i + rev(D_i) & \mbox{if $rev(D_i) \ge D_i$ } \\
\end{matrix}
\right.
\end{eqnarray}
In other words, we compare the digital reversal of a positive integer
$rev(D_i)$ with itself $D_i$: if it is smaller, we subtract the $rev(D_i)$
from $D_i$; otherwise we add the two. Eq.(\ref{eq-map}) has a superficial
similarity with the Collatz map by German mathematician Lothar Collatz in 1937
(see, e.g., 
\citep{lagarias85}): 
\begin{eqnarray}
\label{eq-collatz}
D_{i+1} =
\left\{
\begin{matrix}
D_i/2  & \mbox{if $D_i$ is even} \\
3D_i +1  & \mbox{if $D_i$ is odd } \\
\end{matrix}
\right.
\end{eqnarray}
in the sense that both maps contain a value-increasing and
a value-decreasing component. Therefore, a zigzag trajectory
in the positive integer space is expected for both Eq.(\ref{eq-map})
and Collatz map. At a first glance, Conway's RATS only has
the value-increasing part; however, digital sorting will
decrease the value. So RATS accomplishes both value-increase
and value-decrease operations in one step.

\begin{figure}[th]
\begin{center}
  \begin{turn}{-90}
  \end{turn}
 \includegraphics[width=0.8\textwidth]{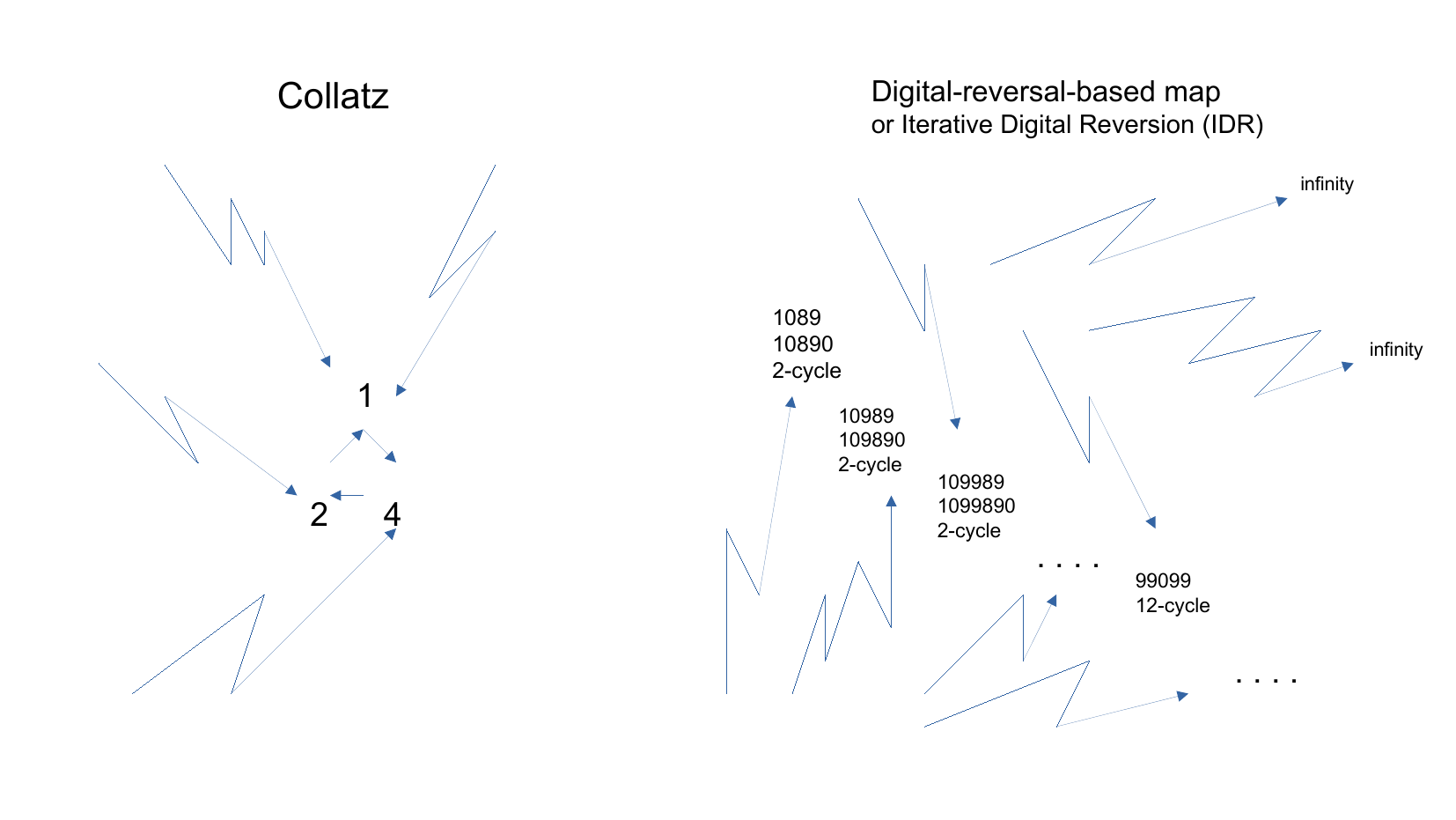}
\end{center}
\caption{
\label{fig0}
Schematic drawing of state space for Collatz map and our digital-reversal
based map in Eq.(\ref{eq-map}). Both have zigzag paths as both maps
consist of two parts, one increasing the state value while another
decreasing it. The Collatz map has a global attractor of a 3-cycle.
Eq.(\ref{eq-map}) map has infinite number of cyclic attractors, with
possible infinite number of cycle lengths, and possible
infinite number of diverging trajectories.
}
\end{figure}

Judging by dynamical behaviors, Eq.(\ref{eq-map}) 
and the Collatz map are completely different. The Collatz conjecture
states that all integers, no matter what the value is,  will 
eventually map to the integer 1, then followed by a 3-cycle
$1 \rightarrow 4 \rightarrow 2 \rightarrow 1$.
Despite a 2021 prize award (https://mathprize.net/posts/collatz-conjecture/),
it is still difficult to prove the  Collatz conjecture, though more people have made
an effort to prove it recently.
Counterexamples have never been found \citep{clay}. It has been proven that
almost all cycle lengths are almost bounded \citep{ttao}, but it has not been proven
that the 3-cycle limiting trajectory is the only limiting set.

Our Eq.(\ref{eq-map}) map has its origin in the ``1089 mathematical trick"
\citep{ball,acheson}. Think of an integer with three digits (e.g., 741);
subtract its digital reversal to get a new integer (741-147=594). Then, add
the new integer with its digital reversal, 594+495=1089.  Think of another
three-digit number, e.g., 981, 981-189=792, 792+297=1089 again! The trick
part is that you can guess the end result (which is 1089) no matter what
the initial three-digit integer is (with the conditions that the first digit
is larger than the last digit and difference is still a 3-digit, not 2-digit, integer). 
If the initial integer has more than three digits, the end results might 
be different from 1089, though still very restricted, such as 10989, 109989, etc.
\citep{webster,AL-1089}.

One might have guessed from the simple outcome from the 1089 trick that
Eq.(\ref{eq-map}) should also lead to a simple limiting attractor, just
like the Collatz map. After extensive computer runs, we have found that
Eq.(\ref{eq-map}) behaves more like RATS than Collatz (a schematic illustration
is shown in Fig.\ref{fig0}), though
more intriguing than RATS in that we found a more restricted number of 
cycle lengths, and it takes 8 iterations to reveal the diverging trajectories as
versus 2 steps for RATS. For details on the digital patterns in
the integers in the limiting dynamics, see \citep{AL-1089} -- here
we focus more on the dynamics aspects of Eq.(\ref{eq-map}).

\begin{figure}[th]
\begin{center}
  \begin{turn}{-90}
  \end{turn}
 \includegraphics[width=0.9\textwidth]{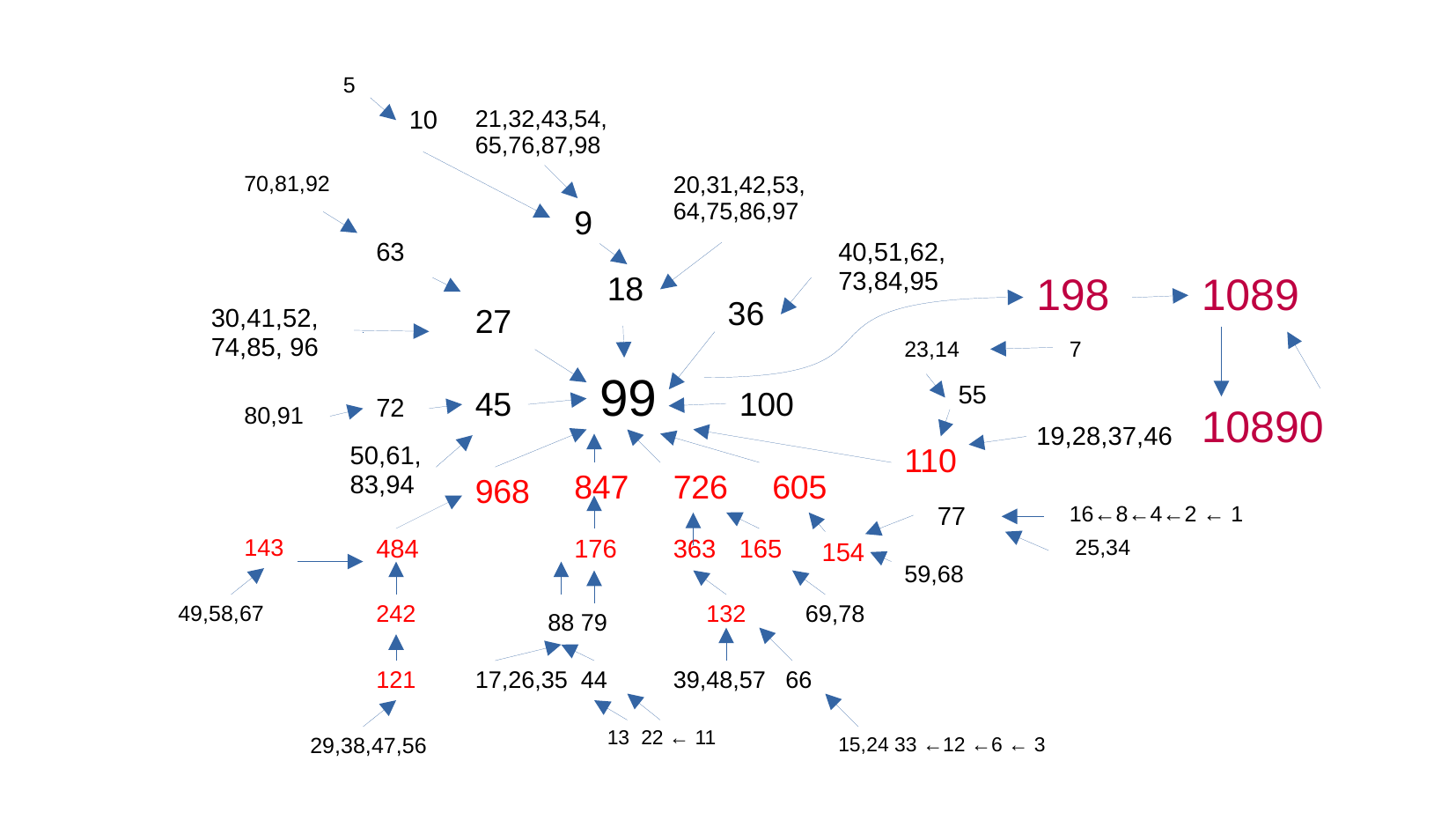}
\end{center}
\caption{
\label{fig1}
The state diagram when the initial state is less than or equal to 100.
All these initial states will be attracted to the state 99 first, then to
198, then to 1089-10890 2-cycle. The states larger than 100 are marked in red.
}
\end{figure}

\section*{Periodicities of 2, 12, 10, and 71}

\indent

\begin{figure}[th]
\begin{center}
  \begin{turn}{-90}
  \end{turn}
 \includegraphics[width=0.8\textwidth]{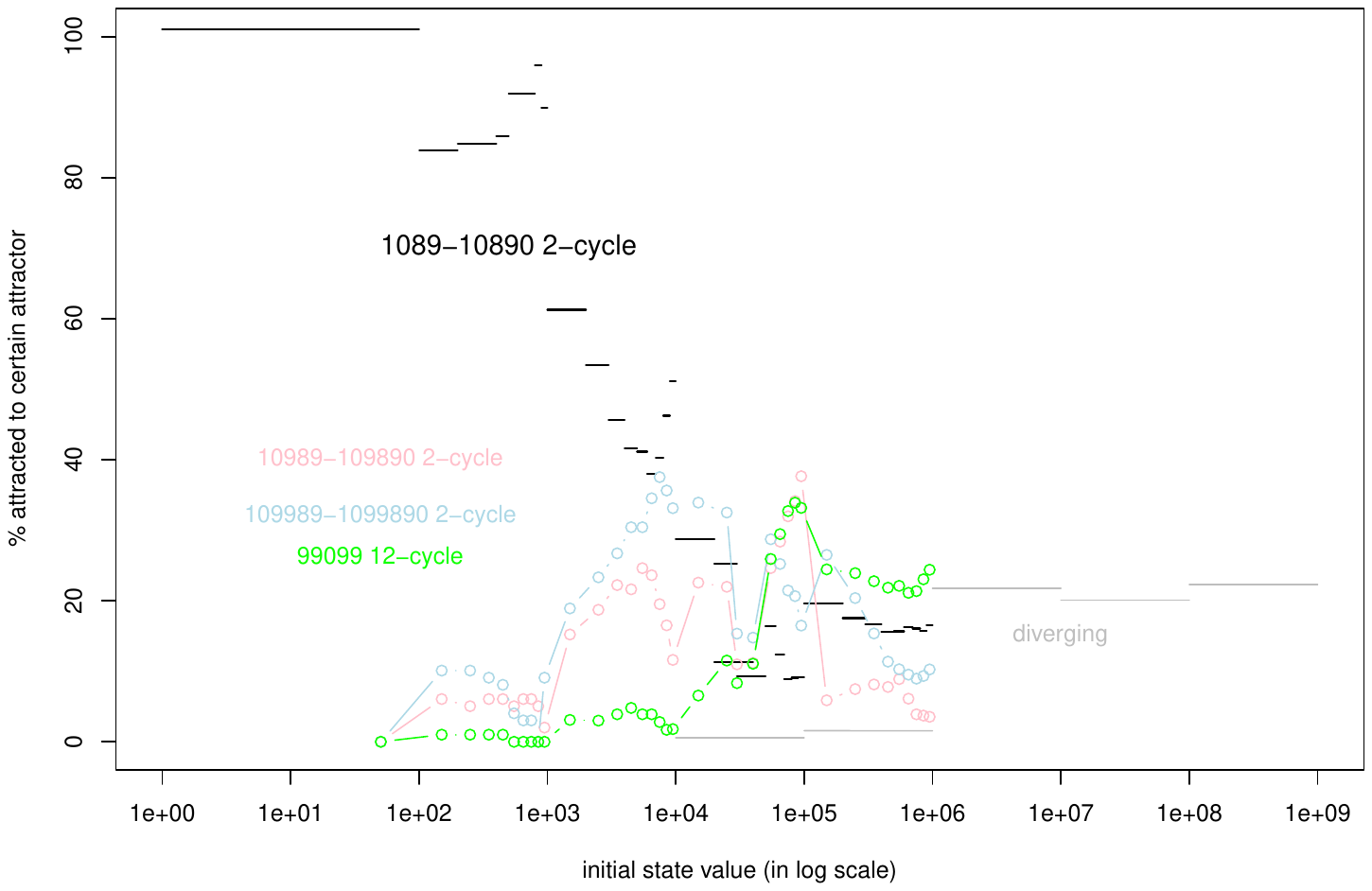}
\end{center}
\caption{
\label{fig2}
Percentage of initial states that are attracted to the 1089-10890 2-cycle
attractor (black), to the 10989-109890 attractor (pink),
to the 109989-1099890 attractor(light blue), and the 99099 21-cycle
attractor (green), as a function of the numeric range of the initial states.
The grey bars, obtained from a larger scale numerical runs,
shows the percentage for diverging trajectories.
}
\end{figure}

When $D_0 \le 100$, we do see a remnant of the 1089 trick as all
eventually map to the two-cycle 1089 $\rightarrow$ 10890 $\rightarrow$ 1089
(see Fig.\ref{fig1}.  Fixed-points limiting dynamics is impossible for
Eq.(\ref{eq-map}) as it would require $rev(D_i)=0$, which violates
our assumption that $D_i$'s are all positive integers.
As $D_0$ increases, other two-cycles appear, such as 10989 $\rightarrow$ 109890 
$\rightarrow$ 10890, which is reached if $D_0=101, 102, 103, 105, 107, 109 \cdots$;
or 109989 $\rightarrow$ 1099890 $\rightarrow$ 109989, if $D_0$=104, 106,
108, 109, 118, 138, 148, 168, 178 $\cdots$.
Fig.\ref{fig2} shows that the percentage of $D_0$'s that are attracted
to the 1089-10890 two-cycle gradually decreases as $D_0$ increases, those attracted
to 10989-109890 or 109989-1099890 gradually increases, then decreases.

In dynamical systems, the limiting sets are also called attractors,
as if the non-attractor transients are ``attracted" to them. It is 
like release a ball from any place in a bowl and the ball will always 
settle at the bottom.  The similar concept of ``attractor" in a 
hill-and-valley landscape is
also used in evolution by Sewall Wright \citep{wright} and in
developmental biology by Conrad Waddington \citep{waddington}.
One may also notice that one element in the two-cycle is 10 times that
of another element. As $D_{i+1}=D_i+rev(D_i)=10D_i$, $rev(D_i)=9D_i$.
Positive integers, whose digital reversal is $k$ times of itself,
are called $k$-palintiple or $k$-palintuple \citep{holt14}.
The 2-cycle limiting set of Eq.(\ref{eq-map}) contains a 9-palintuple.
This also make us wonder if there are other maps whose limiting sets
contain $k$-palintuples where $k \ne 9$.

\begin{figure}[th]
\begin{center}
  \begin{turn}{-90}
  \end{turn}
 \includegraphics[width=0.6\textwidth]{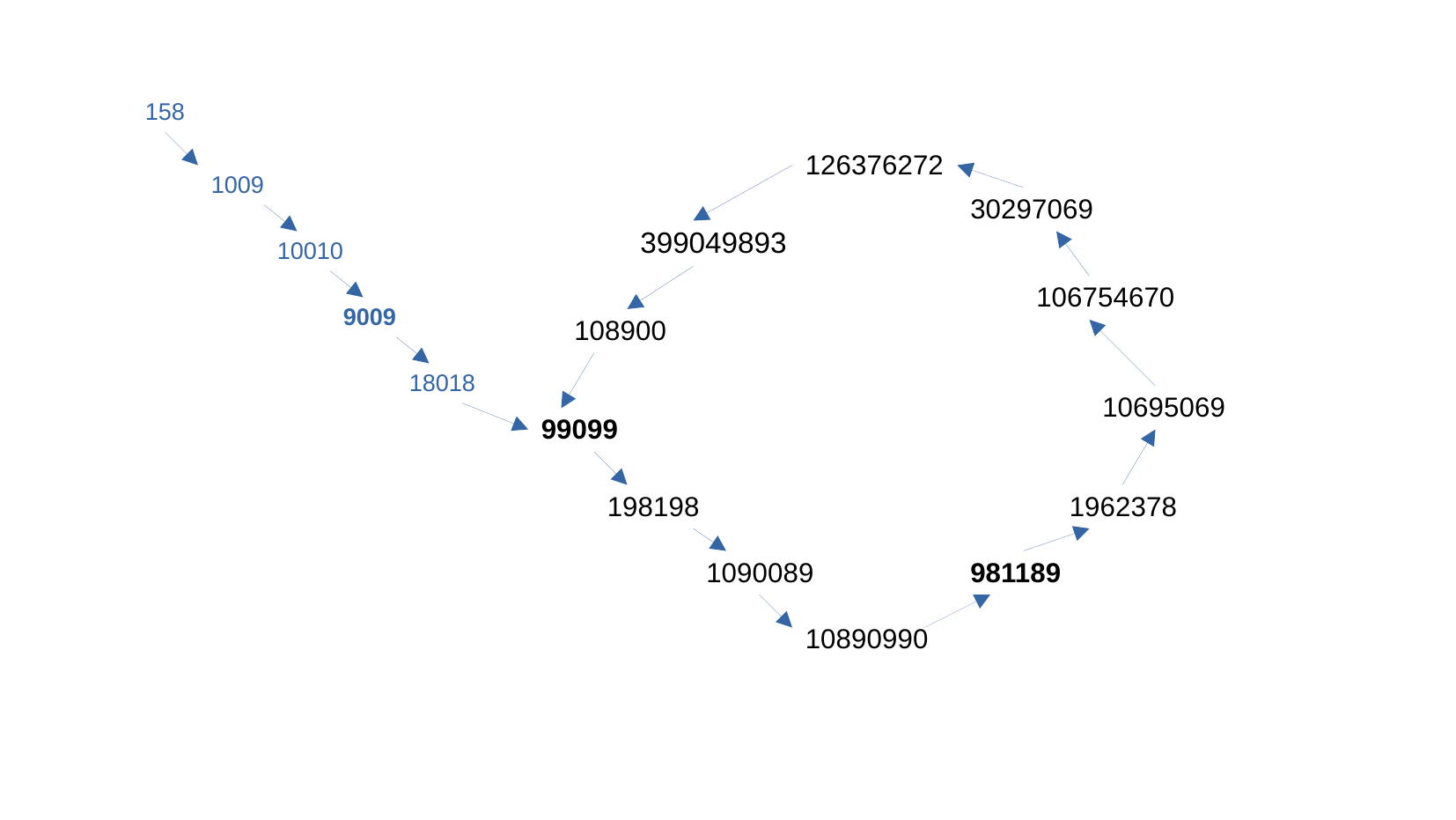}
\end{center}
\caption{
\label{fig3}
The state diagram when the initial state is 158. The trajectory will be attracted
to a 12-cycle. States that are palindromes are marked with bold font.
}
\end{figure}

When $D_0=158$ (or 257, 356, 455, $\cdots$), the limiting dynamics is a 
12-cycle (Fig.\ref{fig3}):
99099 $\rightarrow$
198198 $\rightarrow$
1090089 $\rightarrow$
10890990 $\rightarrow$
981189 $\rightarrow$
1962378$\rightarrow$
10695069 $\rightarrow$
106754670 $\rightarrow$
30297069 $\rightarrow$
126376272 $\rightarrow$
399049893$\rightarrow$
108900 $\rightarrow$ 99099.
We have exhaustively run all initial integers up to $10^9$ (one billion),
and we found 44 2-cycle attractors and 2 12-cycle attractors. No other
cycle lengths were observed.

With $D_0 > 10^9$ we couldn't run the mapping exhaustively, but we randomly
sample $10^7$ (100 millions) integers for each order of magnitude. Other
periodicities have been found. For example, here is a 10-cycle: \\
    1090089109890991089  $\rightarrow$
   10892080098910791990   $\rightarrow$
     972378109902762189   $\rightarrow$
    1953645319804635468   $\rightarrow$
   10599009408940099059   $\rightarrow$
  105698014389430198560   $\rightarrow$
   39806979406019302059   $\rightarrow$
  134827370466517262952   $\rightarrow$
  394090086130590991383   $\rightarrow$
10890991098910900890 $\rightarrow$
1090089109890991089. The integers in this cycle are huge, ranging from
$10^{17}$ to $10^{20}$.

We have also discovered a 71-cycle limiting
dynamics \citep{AL-1089}, where the smallest value is $\sim 2.6 \times 10^{15}$
and the largest $3.6 \times 10^{21}$. The largest $D_0$'s we have attempted
were between $10^{16}$ and $10^{17}$ (100 millions of them randomly sampled),
and no other periodicities were found. Compared to RATS where many periodicity
values have been observed, we only have periodicities of 2, 12, 10, 71 and no more.
It is still an open question on if one can keep finding more, or perhaps
infinitely more, cycle lengths; but we are hopeful.
However, when we run our algorithm for numeral systems beyond decimal, although 
most attractors are still of periodicity 2,
a few other non-standard periodicities have also been found.

\section*{Length-8 diverging trajectories}

\indent

As we mentioned earlier, RATS has diverging trajectories whose digit pattern
(e.g. $12(3)_k444$, $k \ge 2$) repeats after 2 iterations. This was called
length-2 divergence \citep{shattuck}. We will show that Eq.(\ref{eq-map}) has
length-8 diverging trajectories. Starting from $D_0= 109\_99\_98\_9\_0\_000$
where $\_$ simply indicates an empty space making visualization easier, then
after 8 iterations of Eq.(\ref{eq-map}) it maps to 
$109\_9999\_98\_9\_000\_000$ (the overhead dot during
subtraction stage indicates a borrowing from a higher digit, and
a $+$1 during addition stage indicates a contribution of 1 from the
lower digit to the higher one):
\begin{equation}
 \begin{matrix}
  & & & \cdot& \cdot& \\
  & 109 & 99 & 98 & 90 & 000 \\
-)& 000 & 09 & 89 & 99 & 901 \\
 \hline
  & 109 & 90 & 08 & 90 & 099 \\
 \end{matrix}
\end{equation}

\begin{equation}
 \begin{matrix}
  & 109 & 90 & 08 & 90 & 099 \\
+)& 990 & 09 & 80 & 09 & 901 \\
 +1 & & & & +1& \\
 \hline
 1 &099 & 99 & 89 & 00 & 000 \\
 \end{matrix}
\end{equation}

\begin{equation}
 \begin{matrix}
  & & & \cdot& \cdot& \\
  &109 & 999 & 8 & 900 & 000 \\
-)&000 & 009 & 8 & 999 & 901 \\
 \hline
 &109  & 989 & 9 & 900 & 099
 \end{matrix}
\end{equation}

\begin{equation}
 \begin{matrix}
 &109  & 989 & 9 & 900 & 099  \\
+)& 990 & 009 & 9 & 989 & 901 \\
+1 & & +1 & +1 & +1& \\
 \hline
 1&099  &999 &9 & 890 & 000\\
 \end{matrix}
\end{equation}

\begin{equation}
 \begin{matrix}
  & & \cdot& \cdot& \cdot& \\
 &109  &999 &99 & 890 & 000\\
-)&000& 098 & 99 & 999 & 901 \\
 \hline
 & 109 & 900 & 99 & 890 & 099
 \end{matrix}
\end{equation}

\begin{equation}
 \begin{matrix}
   & 109 & 900 & 99 & 890 & 099 \\
+) &990 & 098 & 99 & 009 & 901  \\
 +1 & & +1& & +1& \\
 \hline
 1 &099 & 999 & 98 & 900 & 000
 \end{matrix}
\end{equation}

\begin{equation}
 \begin{matrix}
  & & \cdot & \cdot & \cdot& \\
  &109 &9999 & 9 & 8900 & 000  \\
-)&000 & 0098 & 9 & 9999 & 901 \\
 \hline
 &109 & 9900 & 9 & 8900 & 099
 \end{matrix}
\end{equation}

\begin{equation}
 \begin{matrix}
  &109 & 9900 & 9 & 8900 & 099 \\
+)&990 & 0098 & 9 & 0099 & 901 \\
 +1& & +1& & +1&  \\
 \hline
 1 &099 & 9999 & 8& 9000 & 000
 \end{matrix}
\end{equation}
The pattern 109(9)$_L$ 98 9(0)$_{L-1}$ 000 repeats with $L$ increases by 2
(from $L=2$ to $L=4$), after 8 iterations.

To help to recognize the progression in these 8 iterations, we may mark
the central digit(s) as a signature:
\begin{eqnarray}
\label{eq-infi}
1 & 109 (9)_{L}  \underline{98} 9(0)_{L-1} 000   \rightarrow \nonumber \\
2 & 109 (9)_{L-1}0 \underline{08}  9(0)_{L-1} 099  \rightarrow \nonumber \\
3 & 109 (9)_{L+1} \underline{8} 9(0)_{L} 000   \rightarrow \nonumber \\
4 & 109 (9)_{L-1}89 \underline{9} 9(0)_{L} 099  \rightarrow \nonumber \\
5 & 109 (9)_{L+1}  \underline{99} 89(0)_{L-1} 000  \rightarrow \nonumber \\
6 & 109 (9)_{L-1}00 \underline{99} 89(0)_{L-1} 099 \rightarrow \nonumber \\
7 & 109 (9)_{L+2} \underline{9} 89(0)_{L} 000   \rightarrow \nonumber \\
8 & 109 (9)_{L}00 \underline{9} 89(0)_{L} 099   \rightarrow  \nonumber \\
 & 109 (9)_{L+2} \underline{98} 9(0)_{L+1} 000 
\end{eqnarray}
Then the signature goes like 
98 $\rightarrow$
08 $\rightarrow$
89 $\rightarrow$
9 $\rightarrow$
99 $\rightarrow$
99 $\rightarrow$
9 $\rightarrow$
9 (then back to 98).
To our surprise, the diverging trajectories become more and more common
as $D_0$ increases. Fig.\ref{fig2} (grey) shows the percentage of divergent trajectories
as a function of numerical range of $D_0$'s.  This percentage
shockingly goes up substantially, to around 20\%, when the initial integers
are higher than $10^7$.

\section*{Short transients}

\indent

\begin{figure}[th]
\begin{center}
  \begin{turn}{-90}
  \end{turn}
 \includegraphics[width=0.8\textwidth]{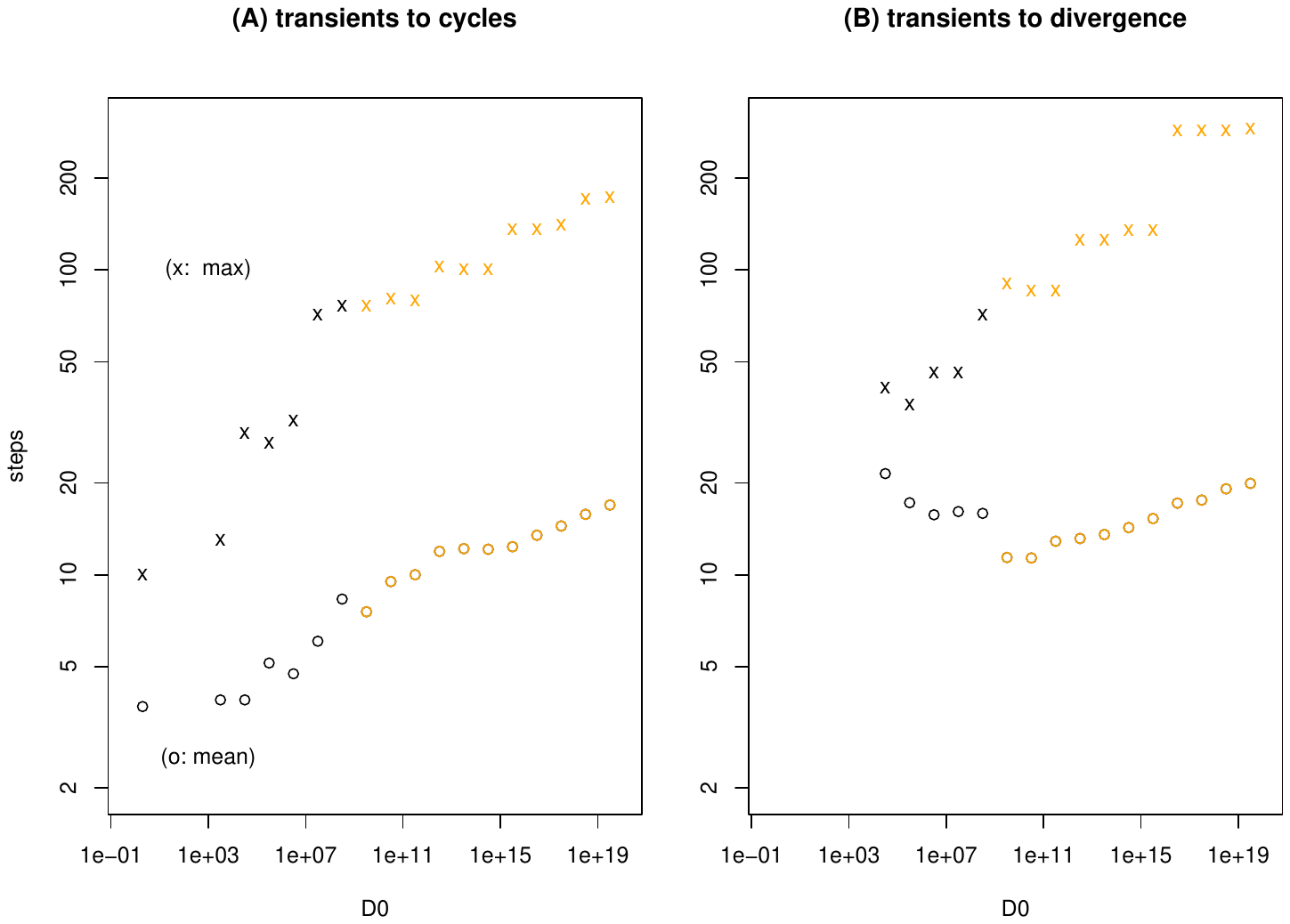}
\end{center}
\caption{
\label{fig5}
Mean (in circles) and maximum (in crosses) of transient steps to
limiting cycles (A) or to diverging trajectories (B), in log-log scale.
Each point represents either an exhaustive numerical run (in black)
or a randomly sampling of 100 millions $D_0$s within an order of magnitude
(e.g. from $10^{11}$ to $10^{12}$). The largest $D_0$s are sampled
in the range of $10^{19}-10^{20}$.
}
\end{figure}

Integers along a trajectory that are not in the attractor themselves,
are called transients. For a given $D_0$, we can measure the transient steps
for both cycles and diverging trajectories. In our numerical runs, we
have observed that the transients are order of magnitude shorter than
one might expect on the ground of randomness.  Fig.\ref{fig5} shows the
mean (circles) and maximum (crosses) of transient steps before reaching
cycles (Fig.\ref{fig5}(A)) or diverging trajectories (Fig.\ref{fig5}(B)).
The points in orange are estimated from 100 millions samples within
a range of order of magnitude for $D_0$ (e.g., from $10^{17}$ to $10^{18}$),
instead of sampling every possible values.

The transients in Fig.\ref{fig5} are only ten steps, on average,
before reaching cyclic limiting sets, and 20 steps reaching diverging 
trajectories. Some initial integers
may have higher transients, taking more than 100 steps to reach
cyclic sets or more than 200 steps reaching diverging sets.
Transient times of any dynamical systems could be affected
by some system parameter (e.g., for cellular automata, a spatially
extended dynamical system,  affected by the number of units \citep{CA110}).
Clearly, the transient time for Eq.(\ref{eq-map}) is affected by
the number of digits in $D_0$.

\vspace{0.5in}

This is the end of our story about map Eq.(\ref{eq-map}). The remaining
extra materials are added in case that could be potentially of interests 
to some readers.

\section*{A map with 4-palintuples in the limiting set}

\indent

As mentioned earlier, we ask if it is possible to design a map whose 
cyclic attractor contains $k$-palintuples ($k \ne 9$ is an integer),
as palintuples have been attracting interests from mathematicians for years
\citep{hardy,sutcliffe,klo,pudwell}.  
We propose this new map from an integer at step $i$, $D_i$, to that at step $i+1$,
$D_{i+1}$, similar to Eq.(\ref{eq-map}) with subtle differences:
\begin{eqnarray}
\label{eq-k4}
D_{i+1} =
\left\{
\begin{matrix}
D_i - rev(D_i) & \mbox{if $rev(D_i) < D_i$} \\
rev\left[ rev \left[ 2 (D_i + rev(D_i)) \right] \right] & \mbox{if $rev(D_i) \ge D_i$ . } \\
\end{matrix}
\right.
\end{eqnarray}
Eq.(\ref{eq-k4}) has 4-palintuples as fixed-point limiting sets.
The purpose of two digit reversals, $rev[rev[]]$ is to get rid of trailing zeros.
The best known 4-palintuple is 2178 \citep{sloane}, as 8712 = 4 $\times$ 2178.
Since (2178+8712) $\times$ 2= 21780, and removing trailing zero is 2178,
then 2178 is a fixed-point of Eq.(\ref{eq-k4}). In fact, Eq.(\ref{eq-k4}) 
is designed with the specific aim of having 4-palintuples as fixed-points, 
because for 4-palintuples, $2(D_i+rev(D_i))= 2(D_i+4D_i)=10D_i$,
which after removing the final 0 is $D_i$ itself.

Preliminary runs of Eq.(\ref{eq-k4}) show that its dynamics is as rich
as Eq.(\ref{eq-map}). When $D_0$ is less than 100, the only limiting dynamics 
is one of the three 4-palintuple fixed-point attractors: 2178, 21978, 219978.
However, with larger $D_0$, both longer cycles and diverging orbits
have been observed.

\section*{Dynamics in the rational number space}

\indent

Consider this simple quadratic map from $X_t$ to $X_{t+1}$, originally defined for 
either real or complex variable $X$'s:
\begin{equation}
\label{eq-x2}
X_{i+1}= X_i^2 +c, 
\end{equation}
If Eq.(\ref{eq-x2}) is limited to rational numbers $X_i$, we can ask the
question about periodic orbits. Eq.(\ref{eq-x2})
has several fixed-points limiting dynamics depending on the value of parameter 
$c$. For $c=0$, $X_0$ =0 is a fixed point; for $c=-2$, both 2 and $-1$ 
are fixed points; for $c=-6$, both 3 and $-2$ are fixed points; etc.  Eq.(\ref{eq-x2}) 
also has a 2-cycle: if $c=-1$, $X_0 = -1 \rightarrow X_1= 0 \rightarrow  X_2=-1$;
and a 3-cycle: if $c=-29/16$, $X_0= -1/4 \rightarrow X_1= -7/4 \rightarrow X_2=5/4
\rightarrow X_3=-1/4$. After that, no higher periodicities have been found. 
 
One may see that the main reason for the difficulty in
finding periodic solutions is because $X_i$ has to be a ratio of two integers
(rational number). Even for the fixed point,
the solutions of $X^2+c=X$, $X=(1 \pm \sqrt{1-4c})/2$, contains square-root,
which usually lead to irrational numbers. Finding rational solutions
for polynomial equations is difficult. Absence of periodic rational
solutions for the map $X_{i+1}=X_i^2+c$ for cycle length 4, 5, 6,
has been proven in \citep{morton,flynn,stoll}. 
There is also a conjecture that no other higher cycle length
solutions exist \citep{MS}. 
The situation for maps on rational
numbers is so different from the maps on positive integers, 
that we are unsure whether the conclusion concerning the lack of high cycles
can be applied to Eq.(\ref{eq-map}).

\section*{Turing halting problem}

\indent

One subtle detail about map Eq.(\ref{eq-map}) not discussed is
why the $rev(D_i)=D_i$ situation belongs to the second line instead of the
first line in Eq.(\ref{eq-map})? The reason is that if we use subtraction
for this situation, the map would stop at $D_i=0$ but we only allow
the dynamics to run in the positive integer space. On the other hand,
we may ask a different question, treating  $rev(D_i)=D_i$ as a singular
point, on whether the map would ever stop/halt: 
\begin{eqnarray}
\label{eq-halt}
D_{i+1} =
\left\{
\begin{matrix}
D_i - rev(D_i) & \mbox{if $rev(D_i) < D_i$} \\
D_i + rev(D_i) & \mbox{if $rev(D_i) > D_i$ } \\
HALT  & \mbox{if $rev(D_i) = D_i$ }
\end{matrix}
\right.
\end{eqnarray}
Eq.(\ref{eq-halt}) may not stop because, as we knew already, the dynamics
run into a periodic or length-8 diverging trajectories. To remove
these possibilities, we ask a new halting problem:
\begin{eqnarray}
\label{eq-halt2}
D_{i+1} =
\left\{
\begin{matrix}
D_i + rev(D_i) & \mbox{if $rev(D_i) \ne D_i$ } \\
HALT  & \mbox{if $rev(D_i) = D_i$ .} 
\end{matrix}
\right.
\end{eqnarray}
In fact, Eq.(\ref{eq-halt2}) is the last item in Richard Guy's book
{\sl Unsolved Problems in Number Theory}:
``it is not known if one repeatedly adds a number to its
reversal, whether a palindrome is always produced" (page 404
of \citep{guy-book}).

What does it have anything to do with the Turing halting problem?
A Turing machine (also called a busy beaver) consists of an infinitely long tape
with binary symbols written on it, a tape reader (and writer), and
the machine is in one of the $n$ states.
Given the machine's state, and given the tape symbol the reading
is facing, an instruction would specify (1) which new symbol to
be  written on the tape to replace the old one; (2) whether the tape reader
head will move to the left or to the  right in one step;
and (3) what is the next machine state. One of the instructions is
to lead the machine to the HALT status.

Many busy beaver machines would never stop. But among those that stop,
the number of steps reaching the HALT status (let's call it transient)
can be quite long. There is a ``busy beaver competition"
to find the maximum number of steps reaching the HALT status \citep{michel22}.
This number increases superfast with the number of states, being
1, 4, 6, 13, 47176870, $> 7.4 \times 10^{36534}$, for busy beaver machines
with $n=$1,2,3,4,5,6 states \citep{aaronson}.  Interestingly,
the 5-state machine with the
maximum 47176870 steps reaching HALT, discovered by Heiner Marxen and
J\"{u}rgen Buntrock \citep{marxen}, is equivalent to the following
Collatz-like map 
\citep{michel22,aaronson}: 
\begin{eqnarray}
\label{eq-bb5}
D_{i+1} =
\left\{
\begin{matrix}
( 5 D_i +18 )/3 & \mbox{if $(D_i) =0$ (mode 3)} \\
(5 D_i +22)/3 & \mbox{if $(D_i) =1$ (mode 3)} \\
HALT  & \mbox{if $(D_i) =2$ (mode 3)} \\
\end{matrix}
\right.
\end{eqnarray}
We wonder: will the transient steps to HALT state in Eq.(\ref{eq-halt2})
increase with the number of digits in $D_0$ in a superfast way?

\section*{Bit reversal in the Fast Fourier Transform}

\indent

Finally we get a chance to discuss digital reversal, whether it
possibly has any real-world applications. What a digital reversal 
accomplishes is to make the least important digit to be the most important one.
Running Eq.(\ref{eq-map}) numerically on a computer will become
more difficult when $D_0$ is large, because the round-off operation
will simply remove the least important digits. Having an arbitrary
precision arithmetic solution is essential to explore the dynamics
of Eq.(\ref{eq-map}) further.

Finding a useful application of digital reversal is hard, but we
found a situation where the reversal of binary sequences play
an important role in an efficient computer algorithm. Suppose
we have these eight 3-bit binary sequences: 000 (=0), 001 (=1), 
010 (=2), 011 (=3), 100 (=4), 101 (=5), 110 (=6), 111 (=7),
where the number after the equal sign is the same value in decimal
representation. Reversing the bits would lead to these mappings:
$1 \leftrightarrow 4$, $3 \leftrightarrow 6$, with 0, 2, 5, 7 map
to themselves. If we change from 0-initiate values $0\dots7$ to
1-initiate values $1\dots8$, the above mappings are
$2 \leftrightarrow 5$, $4 \leftrightarrow 7$, with 1, 3, 6, 8 unchanged.

This bit reversal step (e.g., for 3-bit 1-init indices,
$1,2,3,4,5,6,7,8 \rightarrow 1, 5, 3, 7, 2, 6, 4, 8$)
is an important step in Fast Fourier Transform (FFT),
proposed by James Cooley and John Tukey 
\citep{fft}, 
though the
same idea already appeared in an unpublished manuscript by Carl Friedrich Gauss
in 1805 \citep{heideman}. 
This bit reversal is resulted
from the so called ``butterfly" flow diagram that reshuffles the indices,
and that flow diagram is resulted from splitting the original array
into these with odd and even numbers recursively.

\section*{Acknowledgement}
We would like to thank  Astero Provata for helpful discussions,
and  Oliver Clay for commenting on an early draft.

\normalsize

%
%

\end{document}